\documentclass[12pt]{iopart}

\usepackage{acronym}
\usepackage{siunitx}
\usepackage{caption}
\usepackage{subcaption}
\usepackage{graphicx}
\usepackage{textcomp}
\usepackage{orcidlink}
\usepackage{svg}

\usepackage{lineno}

\begin{document}

\title{Adjustable picometer-stable interferometers for testing space-based gravitational wave detectors}

\author{Marcel Beck$^1$\orcidlink{0009-0007-6116-8578}, Shreevathsa Chalathadka Subrahmanya$^1$\orcidlink{0000-0002-9207-4669} \& Oliver Gerberding$^1$\orcidlink{0000-0001-7740-2698}}

\address{$^1$ Institute of Experimental Physics, University of Hamburg, Luruper Chaussee 149, 22761 Hamburg, Germany}
\ead{marcel.beck@physik.uni-hamburg.de; oliver.gerberding@uni-hamburg.de}
\vspace{10pt}
\begin{indented}
\item[]January 2025 
\end{indented}

\begin{abstract}
Space-based gravitational wave detectors, such as the \ac{LISA}, use picometer-precision laser interferometry to detect gravitational waves at frequencies from \qty{1}{Hz} down to below \qty{0.1}{mHz}.
Laser interferometers used for on-ground prototyping and testing of such instruments are typically constructed by permanently bonding or gluing optics onto an ultra-stable bench made of low-expansion glass ceramic. This design minimizes temperature coupling to length and tilt, which dominates the noise at low frequencies due to finite temperature stability achievable in laboratories and vacuum environments. 
Here, we present the study of an alternative opto-mechanical concept where optical components are placed with adjustable and freely positionable mounts on an ultra-stable bench, while maintaining picometer length stability. With this concept, a given interferometer configuration can be realised very quickly due to a simplified and speed-up assembly process, reducing the realisation time from weeks or months to a matter of hours. We built a corresponding test facility and verified the length stability of our concept by measuring the length change in an optical cavity that was probed with two different locking schemes, heterodyne laser frequency stabilisation and Pound-Drever-Hall locking.
We studied the limitations of both locking schemes and verified that the cavity length noise is below \qty{1}{pm/\sqrt{Hz}} for frequencies down to \qty{3}{mHz}. We thereby demonstrate that our concept can simplify the testing of interferometer configurations and opto-mechanical components and is suitable to realise flexible \acl{OGSE} for space missions that use laser interferometry, such as future space-based gravitational wave detectors and satellite geodesy missions.

\end{abstract}
\vspace{2pc}
\noindent{\it Keywords}: laser interferometry, gravitational wave detection, ultra-stable interferometers
%
%

\section{Introduction}  
The \acf{LISA} is a future space-based gravitational wave observatory that will measure \acp{GW} frequencies from \qty{1}{Hz} down to below \qty{0.1}{mHz}~\cite{LISA_Danzmann2003}, complementary to terrestrial \ac{GW} detectors at audio-band frequencies and pulsar-timings at very low frequencies~\cite{LIGO_2015, Virgo_Acernese2014, Pulsar-Timing_Agazie2023}. \ac{LISA} consists of three spacecrafts, each containing two test masses, forming an equilateral triangle with a distance of \qty{2,5}{million} kilometers between them.
When a gravitational wave passes through the detector, the relative distance between the spacecraft in which the test masses are located changes and is measured by \ac{LISA} using heterodyne laser interferometry.
The required displacement sensitivity within the \ac{LISA} optical system, denoted by $u(f)$, in the sub-Hz regime is specified as
\begin{equation}
    u(f)=1\,\frac{\unit{pm}}{\sqrt{\unit{Hz}}}\times\sqrt{1+\left(\frac{\qty{2}{mHz}}{f}\right)^4} \qquad \left(\qty{e-4}{Hz} < f < \qty{1}{Hz}\right).
\end{equation}

To achieve such low-frequency displacement sensing noise, the optical bench of LISA is going to be constructed by bonding optics onto a low-expansion glass ceramic with a small \ac{CTE}~\cite{Optical_bench_2017}. Similar approaches have been used to built \acf{OGSE} for the mission~\cite{Chwalla2016, Robertson2013, Steier2009}.
The bonding process to realise such interferometers requires precise pre-alignment, extensive planning and preparation, which have motivated the study of alternative approaches to realise such interferometers for ground testing, where, for example, the robustness against mechanical vibrations of a bonded interferometer is not needed. This is further motivated by the need for testing of laser interferometers for other space-based gravitational wave detectors~\cite{Taiji_2021, TianQuin_Luo2016}, for satellite geodesy missions~\cite{Geodesy_Kupriyanov2024, _Geodesy_Danzmann_Sheard_Heinzel_2012}, and for other uses of ultra-stable laser interferometers in areas such as dark matter searches~\cite{ALPS_II, Dark_Matter_optical_cavity} and ultra-stable spectroscopy~\cite{spectroscopy_michelson_XUV, spectroscopy_attosecond}.

We present and demonstrate a concept to realise adjustable laser interferometers satisfying the same low-frequency length stability required for \ac{LISA}, where opto-mechanical components are mounted onto a low-expansion glass ceramic plate with threaded screws and clamps, similar to a typical optical table. The so-realised concept is a further development of a previously demonstrated scheme by Kulkarni \textit{et al.}~\cite{Kulkarni2020} where adjustable mirrors were placed on fixed points of a low-expansion baseplate. Our concept enables the fast realisation of arbitrary interferometer configurations (within the size of the given optical bench), and it allows us to reuse the setup and components for various optical experiments and thus reduce costs, for example, for \ac{OGSE}. We refer to the so-prepared opto-mechanics and the corresponding testing infrastructure as \ac{TAPSI}.

In the initial section of this paper, we describe the opto-mechanical design of our toolset, followed by the optical test facility to shield the interferometer from external noise, especially from temperature fluctuations and mechanical vibrations. To verify the initial stability of our toolset, we set up an optical cavity and measured relative length changes using two different locking schemes, the \ac{HS} and the \ac{PDH} locking. Subsequently, the results and current limitations are discussed, and possible noise sources are addressed. Finally, we provide an outlook for future improvements.
\section{Opto-Mechanical Concept of TAPSI}\label{TAPSI opto-mechanical concept}  
The \acf{TAPSI} combines a glass ceramic optical bench with thermally compensated mirror mounts which are mounted on the bench using low thermal expansion opto-mechanics.
The mirror mounts used are ZeroDrift mirror mounts from Newport with a substantially lower \ac{TTLC} compared to standard (stainless steel) mirror mounts~\cite{ZeroDrift}.
The opto-mechanics are made of Invar while the optical bench is made out of Zerodur 
.
Both materials chosen have a low \ac{CTE}, which is essential to reduce the critical temperature to length (and tilt) coupling at low frequencies.

\begin{figure}[h!]
\centering
\begin{subfigure}{.5\textwidth}
  \centering
  \includegraphics[width=1\linewidth]{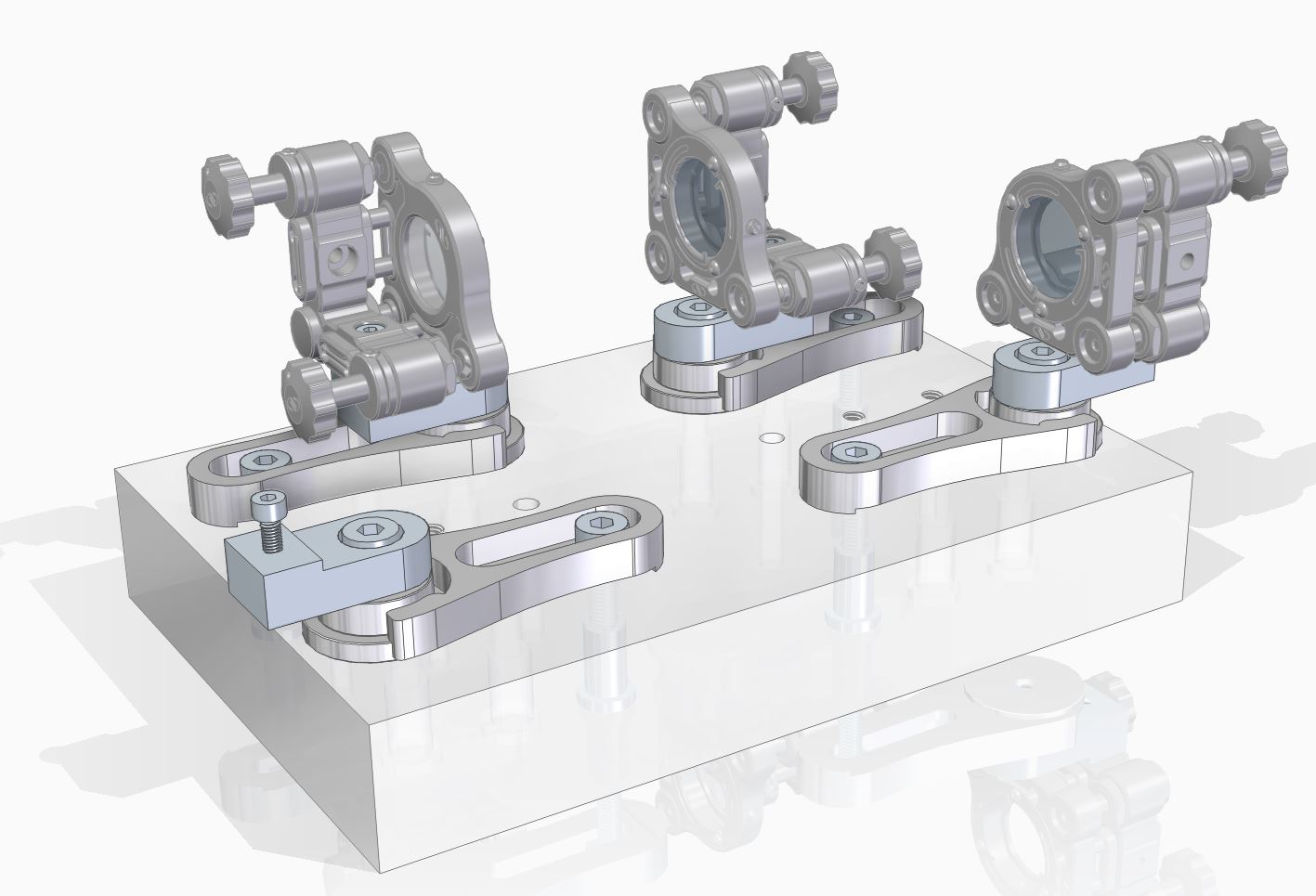}
  \caption{Setup of the prototype cavity}
\end{subfigure}%
\begin{subfigure}{.5\textwidth}
  \centering
  \includegraphics[width=.8\linewidth]{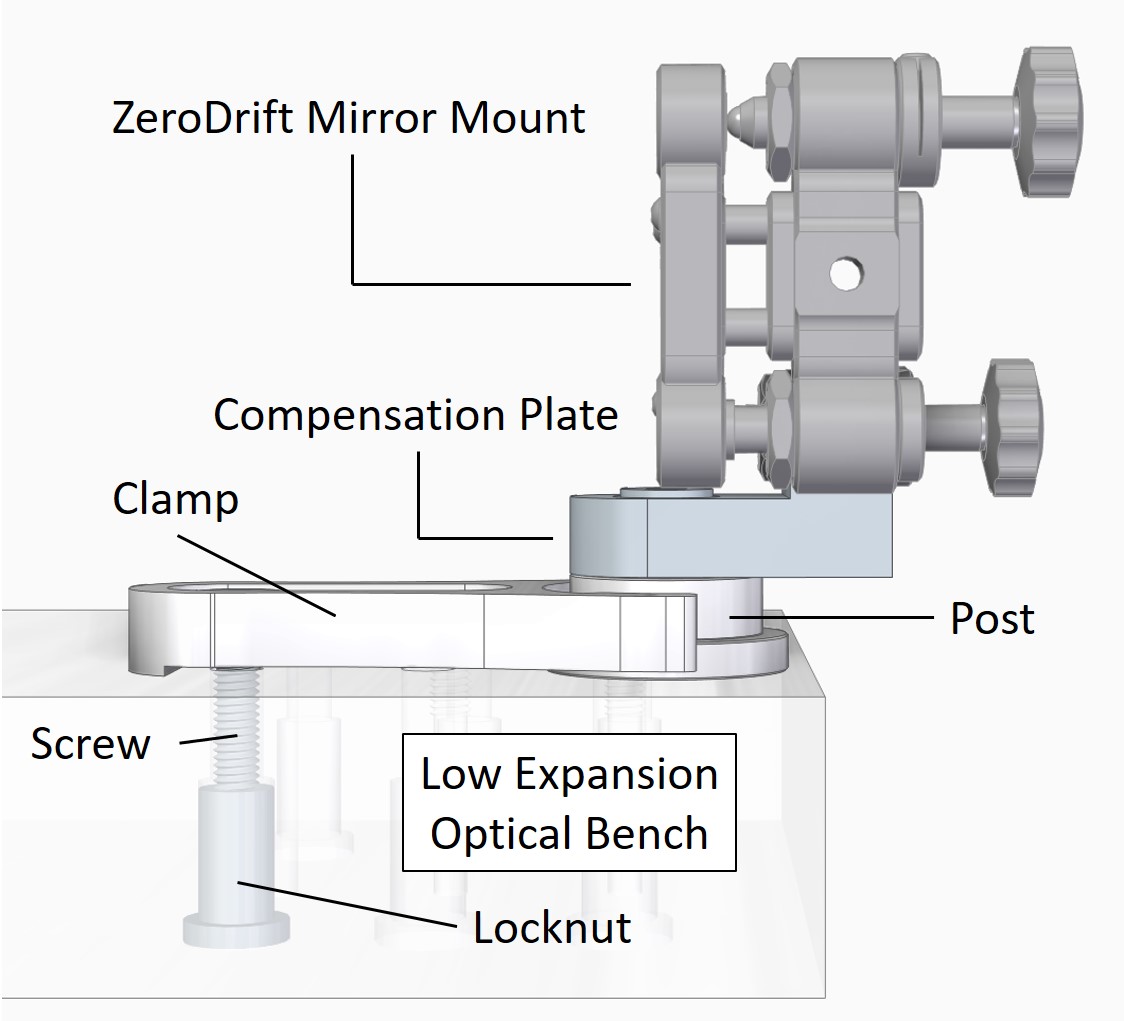}
  \caption{Side view of the mounting concept}
  \label{solid_edge_mounting_concept_labeled}
\end{subfigure}
\caption{CAD model of the \acf{TAPSI} assembly concept and the corresponding setup of the prototype cavity for measuring and verifying the length stability in section~\ref{Laser_Frequency_Locking_section}.}
\label{fig:Prototype cavity.}
\end{figure}

The posts supporting the mirror mounts are fixed onto the bench with clamps and secured with screws and locknuts. 
An additional compensation plate is located between the mount and the post, so that the mounting concept is designed in such a way that the center of the mirror is above the post (see figure~\ref{fig:Prototype cavity.}). This configuration allows for the expansion and contraction of the mirror mount and the Invar compensation plate to cancel each other out. The \ac{CTE} of the thermally compensated mirror mounts was unknown but expected to be small due to the quoted low levels of beam tilts during thermal cycling. Consequently we decided to use Invar for the compensation plate, as it has one of the smallest \acp{CTE}. A future study could investigate the exact \ac{CTE} of the ZeroDrift mirror mounts to further optimise the length and material of the compensation plate.

\section{Test Facility} \label{Optical test facility} 
\begin{figure}[h!]
\centering\includegraphics[width=12cm]{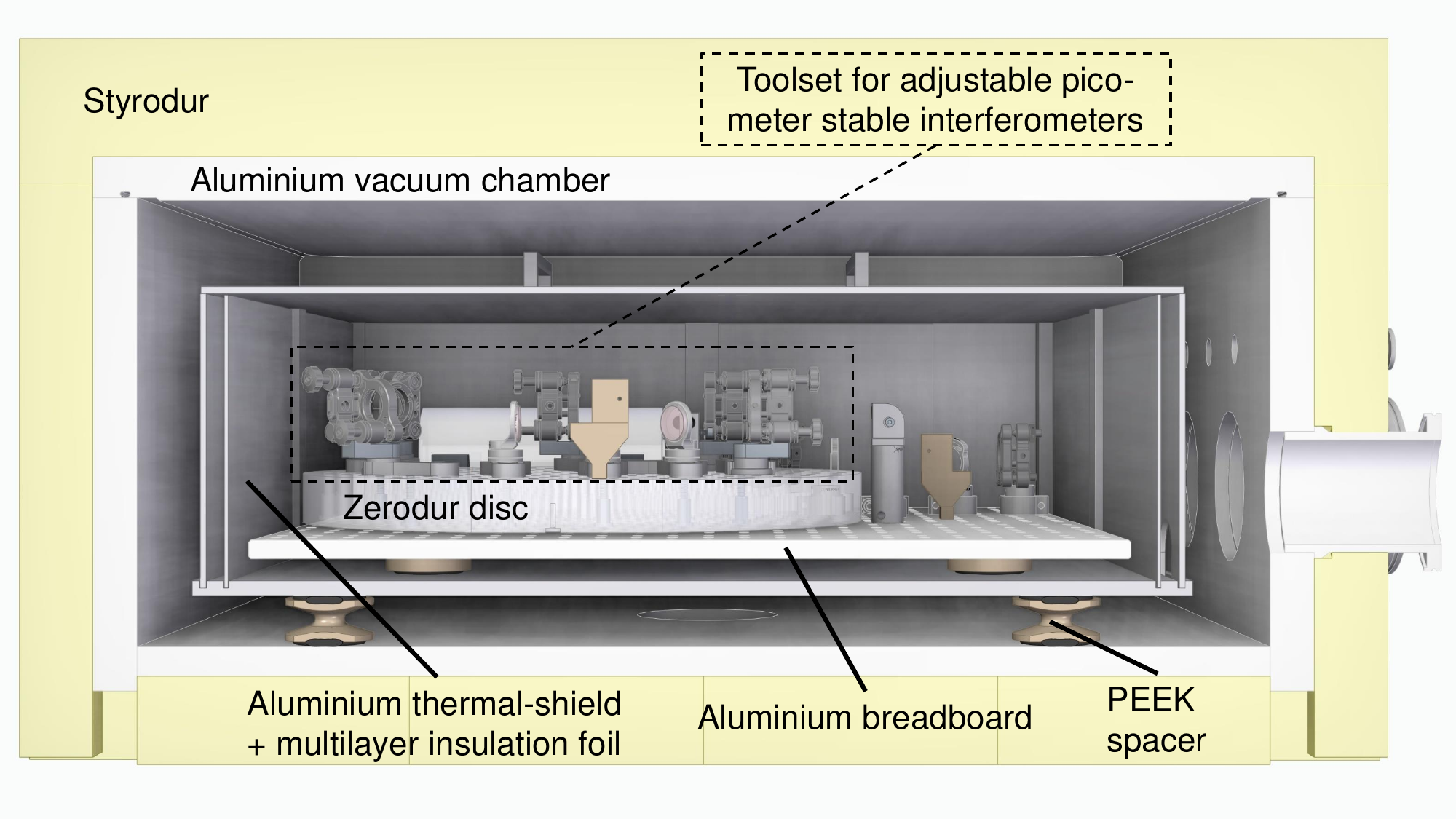}
\caption{Vertical cross-sectional view of the optical test facility with the \acf{TAPSI} inside. A pressure level of \qty{e-3}{mbar} inside the vacuum chamber, multiple thermal isolation layers, and \acs{PEEK} spacers significantly reduce temperature coupling, which is crucial at low frequencies.}
\label{Sectional_view_facility}
\end{figure}

The test facility hosts the optical toolset and consists of a vacuum chamber and multiple thermal shields. The outer thermal shield consists of a Styrodur insulation that houses the vacuum chamber. The double-layer inner thermal shield is made of aluminium and is additionally covered with a multi-layer insulation foil. \acs{PEEK} spacers isolate the chamber, the thermal shield, and the central aluminium breadboard from direct thermal conduction while carrying the corresponding part of the setup. The optical bench, made of Zerodur, rests on the aluminium breadboard by its own weight. A vertical cross-sectional view of the setup is shown in figure~\ref{Sectional_view_facility}.

The test facility is located below a laminar flow box inside a temperature and humidity controlled laboratory.
For the experiments discussed in this paper, we used a scroll pump that provides an equilibrium vacuum of \qty{e-3}{mbar}. The vacuum chamber has an additional flange at the bottom, dedicated to directly host a turbo pump. This provision can be used in future experiments to achieve higher vacuum levels, if needed.

\begin{figure}[h!]
\centering\includegraphics[width=12cm]{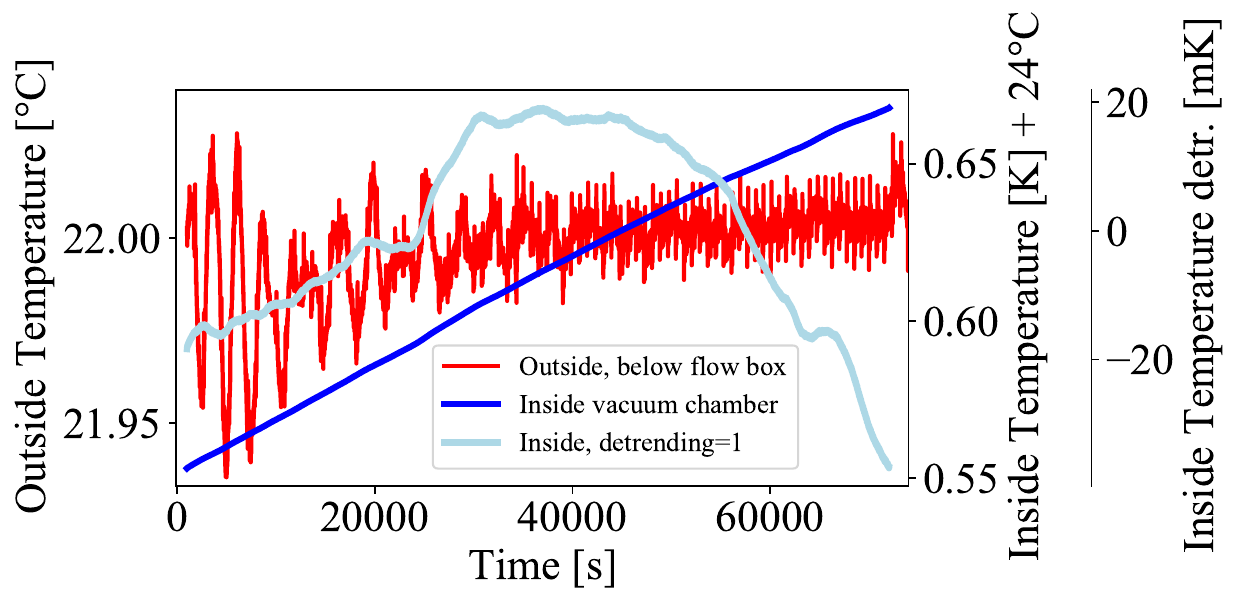}
\caption{Time series of two parallel measurements of temperature inside and outside of the vacuum chamber. The measurement inside the chamber is also shown after subtracting the dominating linear trend caused by the heating of the \acp{PD}.}
\label{PDF_Temperature_time_series}
\end{figure}

\begin{figure}[h!]
\centering \includegraphics[width=12.5cm]{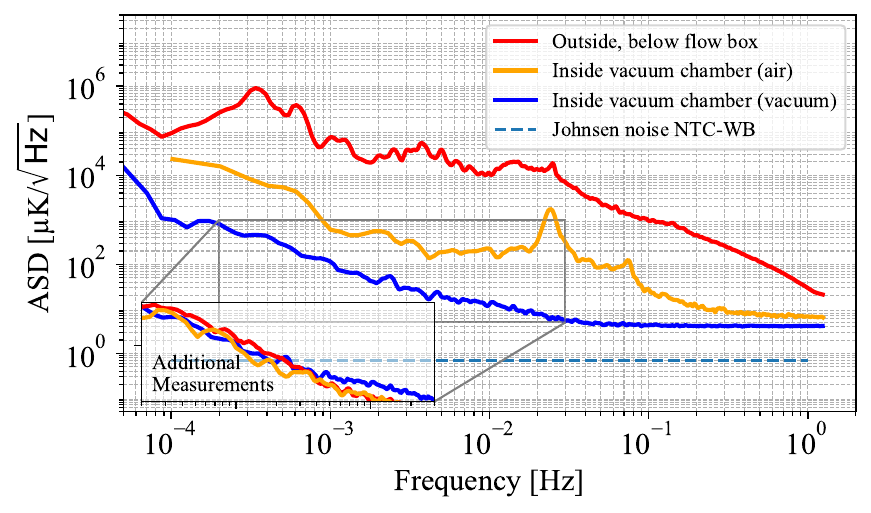}
\caption{\acs{ASD} plot (calculated with \acs{LPSD}~\cite{LPSD_Trbs2006} after removing the linear trend from the measured time series) of the temperature outside and inside the vacuum chamber. The measurements inside the chamber were done before (air) and after reaching vacuum pressure. Additional temperature measurements were taken in conjunction with the stability measurements (figure~\ref{PDF_Freq_ASD_add}) outlined in section~\ref{Results_chapter}. The theoretical sensitivity is given by the Johnsen-Nyquist noise of the Wheatstone bridge resistors.}
\label{PDF_Temperature_ASD_add}
\end{figure}

In order to monitor the temperature fluctuations inside the vacuum chamber, a temperature sensor was developed. The sensor is based on the Wheatstone bridge principle and utilises a \ac{NTC} resistor. The design is based on previous works on temperature sensors for LISA~\cite{TempSensor_RomaDollase2022, TempSensor_Sanjun2007, M_Dehne_2012}. The sensor was located in close proximity to the toolset and the cavity under test in section \ref{Laser_Frequency_Locking_section}. Representative temperature measurements are plotted in figure~\ref{PDF_Temperature_time_series} and the corresponding spectra in figure~\ref{PDF_Temperature_ASD_add}, together with the temperature of the air flow onto the chamber measured with an additional sensor.
Inside the chamber, where the interferometer toolset is placed, a temperature stability of \qty{10}{\micro\kelvin/\sqrt{Hz}} down to \qty{10}{mHz} was achieved.
The theoretical achievable sensitivity is limited by the Johnsen-Nyquist noise of the Wheatstone bridge resistors (\qty{50}{k\Omega}, \qty{85}{k\Omega}) and the \ac{NTC} sensor (\qty{85}{k\Omega}) \cite{Johnson1928, Nyquist1928}.
The time series plot shows no direct temperature coupling between the outside and inside of the chamber, confirming the high degree of thermal isolation of our test facility.
The \acp{PD} inside the chamber were identified as the source of the observed linear increase in temperature. The issue is addressed and discussed in more detail in section~\ref{Results_chapter}.

\newpage
\section{Laser Frequency Locking}\label{Laser_Frequency_Locking_section} 
To verify the length stability of our mounting concept, we locked one of our NPRO-lasers 
(\qty{1064}{nm}) to a prototype cavity with a finesse of \num{627} and a length of \qty{10}{cm} which was set up using the \ac{TAPSI} concept (see figure~\ref{fig:Prototype cavity.}), while the second laser was locked to a reference cavity which was constructed from an \acs{ULE} 
glass, a finesse of \num{10060} and a length of \qty{21}{cm}.
Previous measurements of the reference cavity have demonstrated a stability of \qty{7,5}{fm/\sqrt{Hz}} (\qty{20}{Hz\sqrt{Hz}}) down to \qty{3}{mHz}~\cite{Tröbs:2005}.
The relative length fluctuation between the two cavities is measured by observing the change in beat-frequency of the two locked lasers. This relative length fluctuation $\partial L$ is assumed to be dominated by the length stability of the prototype cavity and is calculated using the following relation
\begin{equation} \label{equ:dL/L=df/f}
\partial L= \frac{\partial \nu}{\nu}\cdot L
\end{equation}
where $L=$ \qty{10}{cm} is the length of the prototype cavity, $\nu=$ \qty{282}{THz} is the laser frequency, and $\partial \nu$ is fluctuation of the beat-frequency. 

We measured the length stability of the prototype cavity using two different locking-schemes, \acf{HS}~\cite{J.Eichholz:15} and the well-known \acf{PDH} locking~\cite{Black2001}. Both techniques are quite similar and lock the laser's frequency to the resonance of the cavity by utilising the interaction phase shift of the cavity's reflected light.
The key difference between \ac{HS} and \ac{PDH} locking is the method to obtain the to-be-stabilised carrier field. While \ac{PDH} locking utilises sidebands which generate an amplitude modulation from a phase modulation with the cavity interaction, \ac{HS} is making use of the already existing beat-note in a heterodyne interferometer. Subsequent demodulation is performed with the sideband modulation frequency in \ac{PDH} locking and with the heterodyne beat-note in \ac{HS}.

\subsection{Monochromatic Beam Reflection and Cavity Interaction Phase} \label{chap:Monochromatic beam reflection and cavity interaction phase}
The electric field of an incident beam can be written as
\begin{equation}
    E_{inc}=E_0 e^{-i\omega t},
\end{equation}
and the reflected beam as
\begin{equation}
    E_{refl}=E_1 e^{-i\omega t},
\end{equation}
where $\omega$ is the (angular) frequency and $E_0$, $E_1$ are the complex amplitudes, describing the relative phase between the two beams.
The ratio of these two beams defines the reflection coefficient
\begin{equation} \label{equ: E_refl / E_in}
    F(\nu)= E_{refl}/E_{inc}=\frac{
    r_1-r_2 e^{i2\pi \frac{\nu}{\text{FSR}}}}
    {1-r_1 r_2 e^{i2\pi \frac{\nu}{\text{FSR}}}},
\end{equation}
where $r$ is the amplitude reflection coefficient of each cavity mirror and $\nu=\omega/2\pi$ is the frequency of the laser light. The \ac{FSR} $=c/2L$ with $L$ as the optical length of the cavity and $c$ as the speed of light.

Far off resonance, the reflection amplitude is approximately equal to one while the phase remains constant ($F(\nu)\approx 1$). In close proximity to the resonance, well within the line width ($\Delta \nu << \delta\nu=\nu - \nu_{res}$), the reflection coefficient is
\begin{equation} \label{equ: refl coefficient}
    F(\delta\nu)\approx \frac{r_1-r_2}{1-r_1 r_2 } - 2\pi i\frac{ r_2 (1-r_1^2)} {(1-r_1 r_2 )^2}\frac{\delta\nu}{\text{FSR}}. 
\end{equation}
For high-finesse cavities 
the imaginary part of the above equation can be rewritten as
\begin{equation} \label{equ:Imaginary part of a high Finesse}
    \mathcal{I}\{F(\delta \nu)\}\approx -\mathcal{G}\frac{\delta\nu}{\Delta \nu}
\end{equation}
with $\mathcal{G}$ summarizing the constants as an optical gain. The resulting equation (\ref{equ:Imaginary part of a high Finesse}) is linear with respect to a frequency change close to resonance.
A slight deviation from the resonance of the reflected beam leads to a phase shift caused by the cavity which is proportional to the change in the length of the cavity itself.

To obtain the phase ($\propto$ length) variation information, the incident beam and the reflected beam from the cavity are measured with a \ac{PD} and subsequently demodulated. This involves mixing and low-pass filtering of the two signals. The resulting so-called error-signal is used to eventually lock the laser's frequency to the resonance of the cavity. Thus, a change of the laser frequency is directly related to a change in the length of the cavity.

At this point, it is imperative to reiterate the distinction between the two locking techniques. While \ac{HS} demodulates the cavity's reflected beam using the already existing heterodyne beat-note, \ac{PDH} locking is using the previously generated sidebands. Each locking technique has its own set of advantages and disadvantages, which are discussed in more detail in the following subsections.

\subsection{Heterodyne Laser Frequency Stabilization} \label{HS_chapter}  
The initial attempt to probe the length stability of the prototype cavity was performed with \acf{HS} which was described and demonstrated before by Eichholz \textit{et. al}~\cite{J.Eichholz:15}. The advantage over \ac{PDH} locking is the simplicity of the locking scheme. \ac{HS} utilises the existing beat-note in a heterodyne interferometer, such as \ac{LISA}, and does not require phase modulators and therefore avoids their noise contribution. 

\begin{figure}[ht!]
\centering\includegraphics[width=12cm]{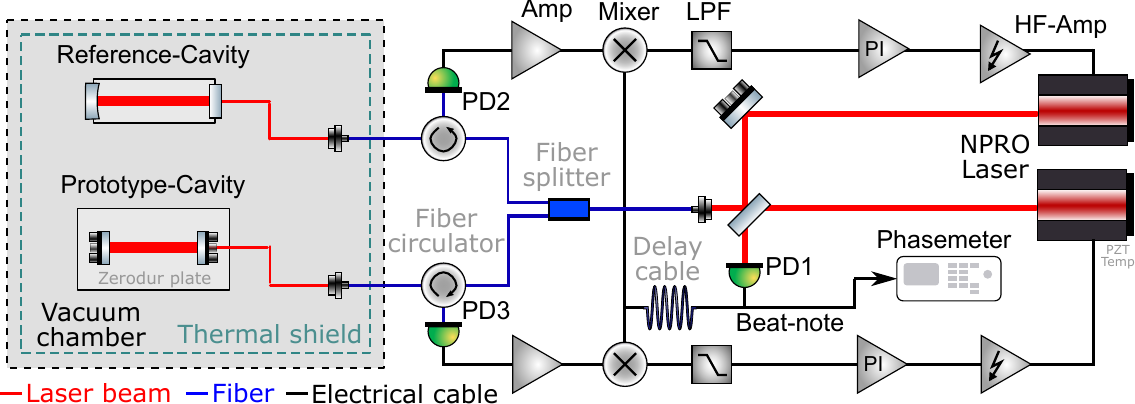}
\caption{The \acl{HS} scheme uses the beat-note from a heterodyne interferometer. The reflected beat-note (as detected by \ac{PD}2 and \ac{PD}3) is then demodulated with the initial beat-note (detected by \ac{PD}1) to extract the cavity interaction phase and create an error-signal. That signal is used to lock each laser to its respective cavity. The measured beat-note by \ac{PD}1 provides the information about the change in relative length of the cavities.}
\label{fig:HS_sketch}
\end{figure}

The \acl{HS} scheme is shown in figure~\ref{fig:HS_sketch}. The interference signal (beat-note) was measured by \ac{PD}1 and also forwarded through a polarization-maintaining, single-mode fiber 
into the vacuum chamber and onto the cavities.
The cavity reflections were detected by \ac{PD}2 and \ac{PD}3, respectively, using fiber-circulators. The measured signals, consisting of the initial beat and the reflected beats from the cavities, were demodulated by analogue mixing and low-pass filtering. With additional phase shifters in front of each mixer, the phase difference was adjusted.
The resulting error-signal was forwarded to a PI-controller 
to lock both lasers to their respective cavities.

The measurement of the \ac{PD}1 beat-note converted to an \ac{ASD} plot is showing the length stability of the prototype cavity realised with our concept \ac{TAPSI} (see figure~\ref{PDF_Freq_ASD_add}). After the first commissioning, we made a number of improvements, including thermal isolation of the optical fibers and electronics of the control loop.

Ultimately, we encountered an implementation limitation that is likely caused by a delay, which we refer to as \textit{delay-beat}. The problem occurs when there is a discrepancy in the path length between the two signals, beginning from the point of interference and ending at the stage of demodulation. This includes the laser's free space propagation, the optical fiber and the electrical cable length.
For example: a linear scan of the laser frequency leads to a linear change in the heterodyne frequency (beat-note).  Due to the differing path lengths, the two signals will arrive at the mixer at different times, with different frequencies, resulting in a phase shift between them. This linear phase shift over time leads to an additional beat, the delay-beat.

Since we used a purely analogue demodulation, a compensation of the delay with an additional electrical cable length was only possible up to a certain degree. Furthermore, long-term frequency drifts will inevitably give rise to issues unless they are actively regulated.
This problem has already been described and overcome in \cite{J.Eichholz:15, Shreevathsa_HS} by introducing a frequency-dependent phase offset in the digital domain. However, the current experimental approach of long-term stability and the noise floor are constrained by the analogue \ac{HS}. The planned transition to digital demodulation will be implemented in future experiments.

\ac{PDH} locking does not encounter the problem of the \textit{delay-beat} because the demodulation frequency, given by the \acp{EOM}, is constant. 

\subsection{\acl{PDH} Locking} \label{PDH-Locking_chapter} 
\begin{figure} [ht!]
\centering
\begin{subfigure}{.5\textwidth}
  \centering
  \includegraphics[width=1.0\linewidth]{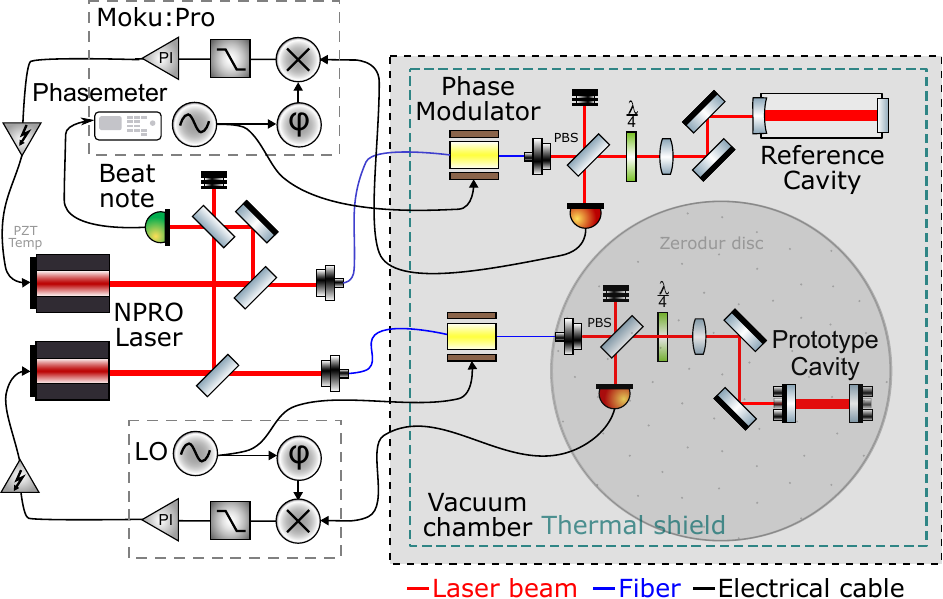}
  \caption{\ac{PDH} locking scheme}
  \label{PDH_sketch}
\end{subfigure}%
\begin{subfigure}{.5\textwidth}
  \centering
  \includegraphics[width=.8\linewidth]{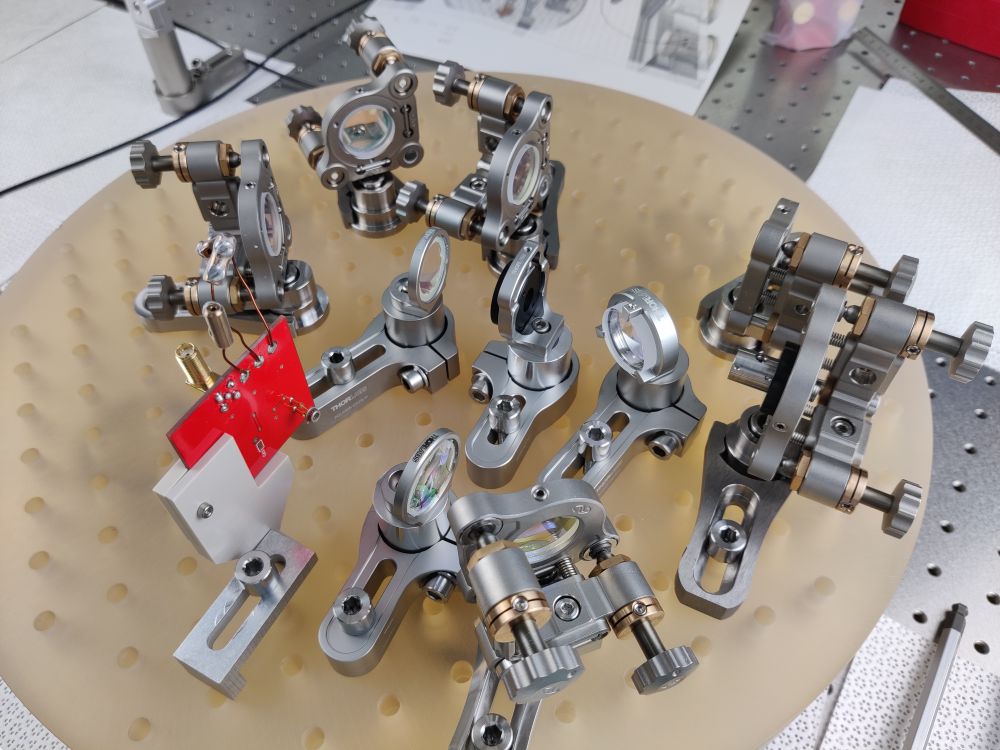}
  \caption{\ac{PDH} locking setup with \ac{TAPSI}}
  \label{Picture_of_TAPSI}
\end{subfigure}
\caption{(a) \ac{PDH} locking scheme with fiber-coupled \acp{EOM} and free space setup inside the vacuum chamber.
The beat-note provides the information about the change in relative length of the cavities and was measured with the phasemeter of the Moku:Pro. (b) Initial placement and alignment of the optics for the \ac{PDH} locking was completed in less than an hour using the \acf{TAPSI}.}
\label{PDH locking lab}
\end{figure}

 As the actual length stability of our prototype cavity, and hence the concept, was not fully probed using the analogue \ac{HS} technique, we switched to the well-known \ac{PDH} locking for that purpose. The final experimental setup is shown in figure~\ref{PDH_sketch}). The optical setup required for \ac{PDH}, including the prototype cavity, were assembled with \ac{TAPSI} (see figure~\ref{Picture_of_TAPSI}). We were able to complete the initial placement and rough alignment of the optics in less than an hour, demonstrating the desired speed for setting up such interferometers.
 
 From the initial commissioning phase, improvements were made to the configuration, such as the transition from fiber circulators to free space, to mitigate the anticipated stray light from the etalon effect of fiber-to-fiber connections.
 
 We also placed the fiber-coupled \acp{EOM} 
 within the vacuum chamber to minimize the \acf{RAM}~\cite{Kokeyama2013_RAM, Korth2016_RAM, Jaatinen2008_RAM}.
 A further challenge associated with \acp{EOM} is the misalignment of the input polarization, which results in an oscillatory rotation of the output polarization. This is subsequently converted to an amplitude modulation, for instance, by a polarization beam splitter. The \acp{EOM} we were using are based on an \ac{APE} waveguide technology whereby only the \ac{TE} mode is guided~\cite{Nekvindov2002_APE_TE_polarization}. Any misalignment of the input polarization results in a static increase in insertion loss. Consequently, our phase modulators served the function of polarization filters as is done in standard practice of \ac{RAM} suppression.
 The signal for the laser light modulation as well as the demodulation, error-signal and PI-controller were done by the multi-instrument mode of one Moku:Pro.

\section{Results} \label{Results_chapter} 
\begin{figure}[ht!]
\centering\includegraphics[width=12.5cm]{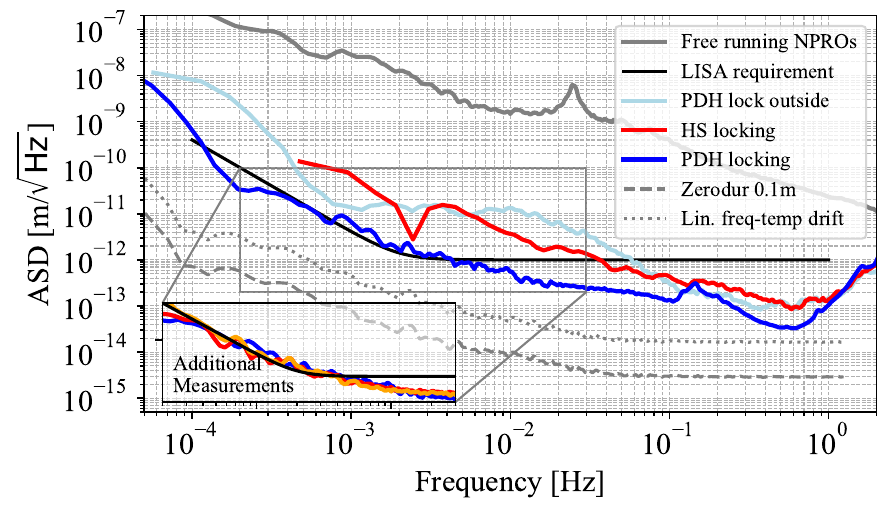}
\caption{The length stability of the prototype cavity was measured with the \acf{HS} and the \acf{PDH} locking techniques. Additional measurements performed with \ac{PDH} locking are plotted in conjunction with temperature measurements from figure~\ref{PDF_Temperature_ASD_add}, highlighting the presence of non-stationary noise.
In addition, the free-running laser noise of a \qty{10}{cm} cavity and the theoretical length deviation of the Zerodur bench in the corresponding environment is presented.}
\label{PDF_Freq_ASD_add}
\end{figure}

The length stability of the adjustable picometer-stable interferometer concept was determined by measuring the length change of a prototype cavity, in comparison to an ultra-stable reference cavity. We have used two different locking schemes, the \acl{HS} (section \ref{HS_chapter}) and \acl{PDH} locking (section \ref{PDH-Locking_chapter}). The results are shown in figure~\ref{PDF_Freq_ASD_add}. Although both results of the locking schemes were improved by thermal isolation, the \ac{HS} was ultimately limited by analogue demodulation. The \ac{PDH} locking scheme, which was setup with the \ac{TAPSI} concept in free space, is below the \ac{LISA} requirements for most frequencies and only slightly above at \qty{0.8}{} to \qty{3}{mHz}, demonstrating the achievable length stability of our concept.

The initial \ac{PDH} configuration, in which the \acp{PD} were placed outside the chamber (see the light blue trace in figure~\ref{PDF_Freq_ASD_add}), revealed the presence of a typical stray light shoulder in the spectrum~\cite{_Stray_light_Fleddermann2018, Stray_light_Sasso2019}. The necessity for additional fiber components in this configuration gave rise to an increased number of fiber-to-fiber connections, leading to the formation of etalons and possibly causing parasitic beams. Furthermore, the initial configuration has incorporated fiber circulators, which have been demonstrated to induce \ac{RAM}~\cite{RAM_paper}.
The best results of the \ac{PDH} locking were achieved by using free space optics as much as possible (see figure \ref{PDH_sketch}).

\begin{figure}[ht!]
\centering\includegraphics[width=13cm]{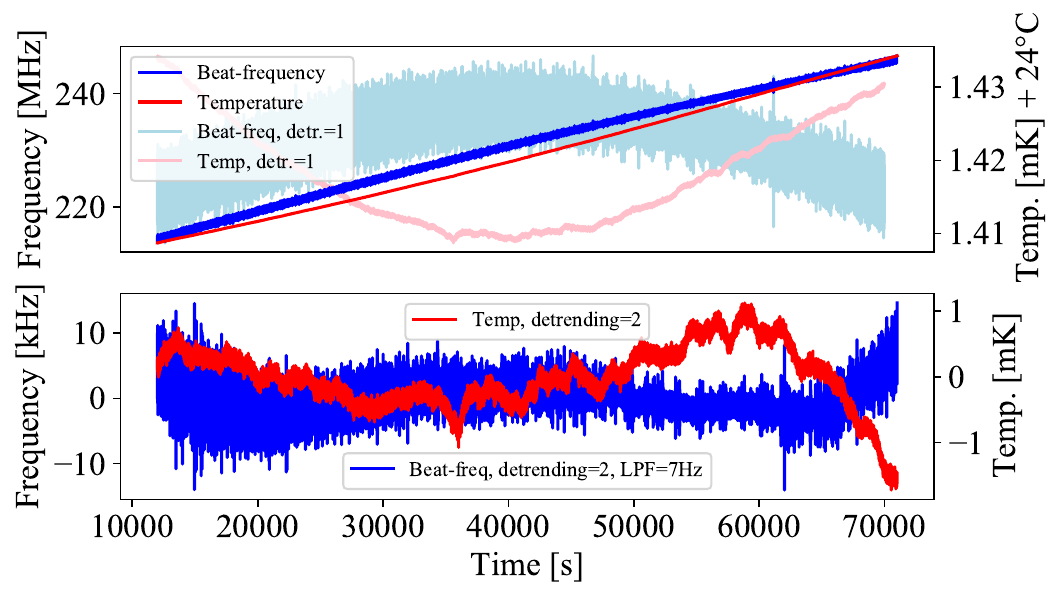}
\caption{The upper subplot shows the time series of parallel measurements of the beat-frequency and the temperature, along with their respective subtraction of the linear drift. In the lower subplot the quadratic drift was removed, while an additional low-pass filter with a cut-off frequency of \qty{7}{Hz} was applied to highlight the presence of non-stationary noise such as parasitic beams.}
\label{PDF_Freq_Temp_time_series}
\end{figure}

Although the \ac{PDH} locking achieves the desired performance, a linear drift in frequency is observed and can be attributed to the linear temperature increase resulting from the \acp{PD} situated in vacuum. The variation in length per degree of temperature change can be determined from figure~\ref{PDF_Freq_Temp_time_series} as
\begin{equation} \label{equ: dL/dT}
    \frac{dL}{dT}= \frac{df}{f_{Laser}}L_{Cavity}\cdot \frac{1}{dT} \approx \qty{4e-8}{m/K}
\end{equation}
where $df$ is the linear rise of the beat-frequency, $f$ the laser frequency, $L_{Cavity}$ the length of the prototype cavity and $dT$ the linear rise of the temperature.
The multiplication of this drift and the temperature inside the chamber (figure~\ref{PDF_Temperature_ASD_add}) provides insight into the linear frequency-temperature dependence. The resulting plot (see figure~\ref{PDF_Freq_ASD_add}) demonstrates that the stability is not limited by this linear temperature coupling, as the projected noise lies below the length stability measurements for all frequencies. Furthermore, we confirmed that there is no discernible correlation or coherence between fluctuations in temperature and length.

The lower subplot of figure~\ref{PDF_Freq_Temp_time_series} shows the beat-frequency time series, after the removal of the quadratic drift and the application of a low-pass filter with a cut-off frequency of \qty{7}{Hz}. The plot reveals the presence of non-stationary noise. The additional length measurements in figure~\ref{PDF_Freq_ASD_add} and the additional temperature measurements from figure~\ref{PDF_Temperature_ASD_add} which were carried out in parallel, highlighting these assumptions. This phenomenon, which may be attributed to factors such as parasitic beams, could potentially explain the observed limitation in our measurement.

The geometrical \acl{TTLC} of our prototype cavity (\qty{0.1}{m}), calculated from the ZeroDrift mirror mounts and our temperature measurement inside the chamber is below \qty{8e-13}{m/K} and can be excluded as a limitation. The theoretical minimum length noise of our length stability experiment is defined by the expansion of the optical bench within the length of the prototype cavity. The \ac{CTE} of the Zerodur material, when multiplied with the temperature noise inside the test facility, yields the minimum length noise and is also plotted in figure~\ref{PDF_Freq_ASD_add}.

\section{Conclusion and Outlook} 
We have developed a \acf{TAPSI} and shown a length stability which satisfies a \qty{1}{pm/\sqrt{Hz}} \ac{LISA} requirement for \ac{OGSE}. The length measurements were performed with \acl{HS} and \ac{PDH} locking in an optical test facility with a temperature stability of \qty{10}{\micro K} down to \qty{10}{mHz}. The current noise limitations of the analogue \ac{HS} is expected to be addressed in future work by moving from analogue to digital demodulation. The \ac{PDH} locking scheme was completely setup with \ac{TAPSI} and the initial alignment was achieved in less than an hour. Subsequent measurements have shown performance satisfying \ac{LISA} requirements and demonstrating the stability of our toolset.
Temperature couplings have not been identified as the predominant limitation; instead, we suspect that other noise sources associated with readout and probably stray light are the main factors limiting performance. The latter limitation could potentially be reduced with a higher finesse in the prototype cavity.

However, further improvements of the test facility are planned for future iterations, including additional thermal shielding and improvements of the temperature sensor to allow monitoring of temperature changes in the lower µK-regime.
In addition, the temperature coupling of the opto-mechanical concept is also to be reduced by replacing the toolset's compensating plate with a \ac{CTE}-compliant material to match the expansion of the ZeroDrift mirror mounts.
An additional improvement, which has already been demonstrated in other \ac{PDH} locking experiments~\cite{Korth2016}, can be achieved through monitoring \ac{RAM} noise, which can be subtracted in post-processing.

In this paper, we demonstrated that glass-ceramic low-expansion optical benches in corresponding environments can be used to realise adjustable picometer-stable interferometers with flexible component placement. Although this concept can be employed to develop \ac{OGSE} for \ac{LISA}, it can also be used for future low-frequency space-based gravitational wave detectors such as Beyond-LISA, Taiji and DECIGO~\cite{Beyond_LISA, Taiji_2021, DECIGO_2021}.

\appendix
\setcounter{secnumdepth}{0}
\section{Data availability statement}
All data that support the findings of this study are included within the article (and any supplementary files).

\section{Acknowledgement}
The authors thank Ortwin Hellmig for the provision of a Zerodur plate to construct the first protoype cavity and for useful discussion, and Christian Darsow-Fromm for the scientific exchange.
We acknowledge funding by the Deutsches Zentrum für Luft- und Raumfahrt (DLR) with funding from the Bundesministerium für Wirtschaft und Klimaschutz under project reference 50OQ2001 and 50OQ2302. The authors also acknowledge support by the Deutsche Forschungsgemeinschaft (DFG, German Research Foundation) under Germany's Excellence Strategy---EXC 2121 ``Quantum Universe''---390833306.

\section{List of Abbreviations}
\begin{acronym}
\acro{TAPSI}{toolset for adjustable picometer-stable interferometers}
\acro{LISA}{Laser Interferometer Space Antenna}
\acro{GW}{gravitational wave}
\acro{OGSE}{optical ground support equipment}
\acro{HS}{Heterodyne laser frequency Stabilization}
\acro{PDH}{Pound-Drever-Hall}
\acro{CTE}{coefficient of thermal expansion}
\acro{TTLC}{tilt to length coupling}
\acro{ASD}{amplitude spectral density}
\acro{FSR}{free spectral range}
\acro{EOM}{electro-optic modulator}
\acro{RAM}{residual amplitude modulation}
\acro{PD}{photo-detector}
\acro{ASD}{amplitude spectral density}
\acro{APE}{annealed proton-exchanged}
\acro{TE}{transverse electrical}
\acro{PEEK}{Polyether ether ketone}
\acro{NTC}{negative temperature coefficient}
\acro{LPSD}{linear power spectral density}

\end{acronym}

\section{References}
\bibliographystyle{plain}
\bibliography{Bib}

\end{document}